\documentclass[onecolumn,epjc3]{svjour3}  

\smartqed  
\RequirePackage{graphicx}
\RequirePackage[numbers,sort&compress]{natbib}
\usepackage{graphicx}
\usepackage{bm}
\usepackage{amsmath}
\usepackage{amsfonts}
\usepackage{amssymb}
\usepackage{multirow}
\usepackage[colorlinks=True, linkcolor=red, citecolor=red, urlcolor=blue]{hyperref}
\usepackage{adjustbox}
\usepackage{array}
\usepackage{acro}
\usepackage[utf8]{inputenc}
\usepackage{CJKutf8}

\RequirePackage{fix-cm}

\def\({\left(}
\def\){\right)}
\def\[{\left[}
\def\]{\right]}

\newcommand{\be}{{\begin{eqnarray}}}
\newcommand{\ee}{{\end{eqnarray}}}

\newcommand{\Gpcyr}{{\rm Gpc^{-3}~yr^{-1}}}

\DeclareAcronym{LIGO}{
  short = \emph{LIGO} ,
  long = Laser Interferometer Gravitational-Wave Observatory ,
  short-plural = ,
}
\DeclareAcronym{LVC}{
  short = LVC ,
  long = {LIGO Scientific and Virgo Collaborations} ,
  short-plural = ,
}
\DeclareAcronym{LVK}{
  short = LVK ,
  long = {Advanced LIGO, Virgo and KAGRA Collaborations} ,
  short-plural = ,
}
\DeclareAcronym{GW}{
  short = GW ,
  long = gravitational wave ,
  short-plural = s ,
}
\DeclareAcronym{SGWB}{
  short = SGWB ,
  long = stochastic gravitational-wave background ,
  short-plural = s ,
}
\DeclareAcronym{CBC}{
  short = CBC ,
  long = compact binary coalescence ,
  short-plural = s ,
}
\DeclareAcronym{NSBH}{
  short = NSBH ,
  long = neutron star-black hole ,
  short-plural = s ,
}
\DeclareAcronym{BBH}{
  short = BBH ,
  long = binary black hole ,
  short-plural = s ,
}
\DeclareAcronym{PBH}{
  short = PBH ,
  long = primordial black hole ,
  short-plural = s ,
}
\DeclareAcronym{SNR}{
  short = SNR ,
  long = signal-to-noise ratio ,
  short-plural = s ,
}
\DeclareAcronym{IMRPPv2}{
  short = ,
  long = {\normalsize IMRP}{\footnotesize HENOM}{\normalsize P}v2 ,
  short-plural = ,
}
\DeclareAcronym{PTA}{
  short = PTA ,
  long = pulsar timing array ,
  short-plural = s ,
}
\DeclareAcronym{SFR}{
  short = SFR ,
  long = star formation rate ,
  short-plural =  ,
}
\DeclareAcronym{FRW}{
  short = FRW ,
  long = Friedman-Robertson-Walker ,
  short-plural =  ,
}
\DeclareAcronym{IMR}{
  short = IMR ,
  long = inspiral-merger-ringdown ,
  short-plural =  ,
}
\DeclareAcronym{LISA}{
	short = LISA ,
	long  = Laser Interferometer Space Antenna
}
\DeclareAcronym{ET}{
	short = ET ,
	long  = Einstein Telescope
}
\DeclareAcronym{CE}{
	short = CE ,
	long  = Cosmic Explorer
}
\DeclareAcronym{ABH}{
	short = ABH ,
	long  = astrophysical black hole
}
\journalname{Eur. Phys. J. C}
\begin{document}

\title{GW200105 and GW200115 are compatible with a scenario of primordial black hole binary coalescences
}


\author{Sai Wang\thanksref{addr1,addr2}
        \and
        Zhi-Chao Zhao\thanksref{e1,addr3} 
}

\thankstext{e1}{e-mail: zhaozc@bnu.edu.cn (Corresponding author)}


\institute{Theoretical Physics Division, Institute of High Energy Physics, Chinese Academy of Sciences, Beijing 100049, P. R. China\label{addr1}  \and School of Physical Sciences, University of Chinese Academy of Sciences, Beijing 100049, P. R. China\label{addr2} \and Department of Astronomy, Beijing Normal University, Beijing 100875, P. R. China\label{addr3}
}

\date{Received: date / Accepted: date}

\maketitle

\begin{abstract}
Two gravitational wave events, i.e. GW200105 and GW200115, were observed by the Advanced LIGO and Virgo detectors recently. In this work, we show that they can be explained by a scenario of primordial black hole binaries that are formed in the early Universe. The merger rate predicted by such a scenario could be consistent with the one estimated from LIGO and Virgo, even if primordial black holes constitute a fraction of cold dark matter. The required abundance of primordial black holes is compatible with the existing upper limits from microlensing, caustic crossing and cosmic microwave background observations. 
\keywords{primordial black hole \and gravitational wave}
\end{abstract}

\section{Introduction}

Based on a second half of the third observing run, the \ac{LVC} \cite{LIGOScientific:2021qlt} reported two gravitational wave events, namely GW200105 and GW200115, which are compatible with \ac{NSBH} binaries. 
At $90\%$ confidence level, the primary components were found to be black holes with masses of $8.9_{-1.5}^{+1.2}M_\odot$ and $5.7_{-2.1}^{+1.8}M_\odot$, respectively, while the secondary ones $1.9_{-0.2}^{+0.3}M_\odot$ and $1.5_{-0.3}^{+0.7}M_\odot$, respectively. 
Compared with the maximal mass of neutron stars, both of the secondaries are compatible with neutron stars with a probability of $\sim90\%$. 
When the events were assumed to be representative of the entire \ac{NSBH} population, the merger rate densities of GW200105 and GW200115 were inferred to be $16_{-14}^{+38}~\Gpcyr$ and $36_{-30}^{+82}~\Gpcyr$, respectively. 
When a broader distribution of component masses was assumed, they became $130_{-69}^{+112}~\Gpcyr$. 
The above rates were found to be consistent with the model predictions of \ac{NSBH} formation in isolated binaries or young star clusters. 
For either of the two events, however, \ac{LVC} found no evidence of tides or tidal disruption meanwhile identified no electromagnetic counterparts. 
Therefore, there is no direct evidence that the secondaries are neutron stars. 

In this work, we show that a scenario of \ac{PBH} binary coalescences can explain the origin of GW200105 and GW200115. 
Gravitational collapse of enhanced overdensities could produce \acp{PBH} in the early Universe \cite{Hawking:1971ei,Carr:1974nx,GarciaBellido:1996qt,Clesse:2015wea,Dolgov:2013lba,Harada:2013epa,Harada:2016mhb,Khlopov:2008qy,Belotsky:2014kca,Ketov:2019mfc,Zhou:2020kkf}. 
\acp{PBH} can form binaries through several formation channels \cite{Bird:2016dcv,Nakamura:1997sm}. 
Amongst these channels, the most efficient one was originally proposed in Ref.~\cite{Sasaki:2016jop,Nakamura:1997sm}, and recently extensively studied in Refs.~\cite{Raidal:2017mfl,Ali-Haimoud:2017rtz,Chen:2018czv} by considering such as a generic mass distribution of \acp{PBH} and the tidal forces from all the neighboring \acp{PBH} and linear density perturbations. 
To determine if GW200105 and GW200115 are primordial, we have to check whether their merger rate densities can be accounted for with the corresponding \ac{PBH} abundance allowed by the existing upper limits (see review in Ref.~\cite{Carr:2020gox} and references therein). 
In the following, we will compute the merger rate density of \acp{PBH} in two specific models by following Ref.~\cite{DeLuca:2020sae}. 
The first one assumes that the mass function is determined by GW200105 and GW200115. 
The second one assumes that all the black hole binary coalescences observed by \ac{LVC} are primordial. 
The predicted rates will be compared with the estimated ones from \ac{LVC}. 

The rest of the paper is arranged as follows.
In Sec.~\ref{sec:mr}, we show the formula for the merger rate of \ac{PBH} binaries. 
In Sec.~\ref{sec:mf}, we briefly review the log-normal mass function of \acp{PBH}. 
Our results and conclusions are shown in Sec.~\ref{sec:re} and Sec.~\ref{sec:cd}, respectively.

\section{Merger rate of PBH binaries}\label{sec:mr}

We consider the formation channel of \ac{PBH} binaries in the early Universe \cite{Nakamura:1997sm}. 
It is known that this channel makes a dominant contribution to the \ac{PBH} merger rate \cite{Ali-Haimoud:2017rtz}. 
This means neglects of the binary formation mechanisms, such as dynamical captures and three-body interactions, in the late Universe \cite{Kritos:2020wcl}. 
The late-Universe formation mechanism was originally proposed in Ref.~\cite{Nakamura:1997sm} and revisited recently in Ref.~\cite{Sasaki:2016jop}. 
Two neighboring \acp{PBH} could form a binary due to torque from a third neighboring \ac{PBH}, and merge with each other within the age of the Universe due to the energy loss via gravitational emissions. 
However, the mass distribution of \acp{PBH} was assumed to be monochromatic in the original literature. 
More recently, the merger rate of \acp{PBH} with an extended mass function was studied in Refs.~\cite{Raidal:2017mfl,Ali-Haimoud:2017rtz,Chen:2018czv}. 
To be specific, Ref.~\cite{Raidal:2017mfl} took into account the tidal force from a \ac{PBH} which is closest to the center of mass of \ac{PBH} binary. 
Ref.~\cite{Ali-Haimoud:2017rtz} assumed a flat mass function of \acp{PBH} within a narrow mass range. 
To generalize the above two works, Ref.~\cite{Chen:2018czv} took into account the tidal forces from all neighboring \acp{PBH} and linear density perturbations, and meanwhile considered a generic mass function of \acp{PBH}. 
Therefore, we follow Ref.~\cite{Chen:2018czv} to compute the merger rate of \ac{PBH} binaries with a log-normal mass distribution. 

For \acp{PBH} within mass intervals of $(m_i,m_i+dm_i)$ and $(m_j,m_j+dm_j)$, the merger rate per unit volume within temporal interval $(t,t+dt)$ is defined as $R(t)=\mathcal{R}(m_i,m_j,t)dm_{i}dm_{j}$ in units of $\Gpcyr$.
The comoving merger rate density is obtained as follows \cite{Chen:2018czv}
\begin{align}
\label{eq:mr}
\mathcal{R}(m_i,m_j,t)=~&3.9\times10^{6}\times\left(\frac{t}{\tau}\right)^{-\frac{34}{37}}f^{2}\left(f^{2}+\sigma_{\mathrm{eq}}^{2}\right)^{-\frac{21}{74}} \mathrm{min}\left(\frac{P(m_i)}{m_i},\frac{P(m_j)}{m_j}\right) \nonumber\\
&\times\left(\frac{P(m_i)}{m_i}+\frac{P(m_j)}{m_j}\right) \left(m_i m_j\right)^{\frac{3}{37}}\left(m_i+m_j\right)^{\frac{36}{37}} \ ,
\end{align}
where $f=\Omega_{\mathrm{pbh}}/\Omega_{\mathrm{m}}$ is the total fraction of non-relativistic matter in \acp{PBH}, $\sigma_{\mathrm{eq}}\approx 0.005$ is the variance of overdensities of the rest of dark matter on scales of order $\mathcal{O}(10^{-2}-10^{5})M_\odot$ at equality \cite{Ali-Haimoud:2017rtz}, $\mathrm{min}(x_1,x_2)$ selects the minimal value between $x_1$ and $x_2$, $P(m)$ is the mass function of \acp{PBH}, $t$ and $\tau$ denote the cosmic time and the present age of the universe, respectively.
The abundance of \acp{PBH} in dark matter is $f_{\mathrm{pbh}}= \Omega_{\mathrm{pbh}}/\Omega_{\mathrm{dm}} \simeq f \Omega_{\mathrm{m}}/\Omega_{\mathrm{dm}}$.
Here, $ \Omega_{\mathrm{pbh}}$, $\Omega_{\mathrm{dm}}$ and $\Omega_{\mathrm{m}}$ denote the energy density fractions of \acp{PBH}, dark matter and non-relativistic matter, respectively, in the critical energy density of the Universe at present. 
In addition, we would not consider the redshift evolution of the merger rate density, since only the low-redshift events are detectable for \ac{LVC}.
Throughout this work, we fix all cosmological parameters to be the best-fit values from the Planck 2018 results \cite{Aghanim:2018eyx}.

\section{Mass function of PBHs}\label{sec:mf}

The mass function of \acp{PBH} can be determined by the production mechanisms of \acp{PBH} in the early Universe. 
The most likely one is based on the gravitational collapse of overdensities in the radiation dominated epoch of the Universe \cite{Hawking:1971ei,Carr:1974nx,GarciaBellido:1996qt,Clesse:2015wea,Dolgov:2013lba,Harada:2013epa,Harada:2016mhb}. 
Therefore, the mass function of \acp{PBH} depends on the properties of primordial curvature perturbations. 
It is usually parametrized to be a log-normal function as follows
\be
P(m)=\frac{1}{\sqrt{2\pi}\sigma m}\mathrm{exp}\(-\frac{\ln^{2}(m/m_{\ast})}{2\sigma^{2}}\)\ ,
\ee
where $m_{\ast}$ and $\sigma$ stand for the center mass and width, respectively \cite{Chen:2018czv,Carr:2017jsz,DeLuca:2020sae}. 
In this work, we consider two different methods to determine values of $m_{\ast}$ and $\sigma$. 
First, we assume that all observed black hole binaries are primordial. 
The results have been shown to be $m_{\ast}=19M_\odot$ and $\sigma=0.97$ \cite{DeLuca:2020sae}, which will be used in this work. 
There are also alternative choices of mass function \cite{Chen:2018czv}, which would not significantly change our results of the abundance of \acp{PBH}. 
Second, we assume that only GW200105 and GW200115 are of primordial origin and they determine the mass function of \acp{PBH}. 
Typically, we choose one half of the total mass to determine the value of $m_\ast$, namely $m_\ast=5.4M_\odot$ for GW200105 and $m_\ast=3.6M_\odot$ for GW200115. 
Meanwhile, we set the width to be $\sigma=m_\ast/10M_\odot$, implying $\sigma=0.54$ for GW200105 and $\sigma=0.36$ for GW200115. 
In such a case, slightly different choices would change our predictions on the merger rate density by a factor of $\mathcal{O}(1)$. 
Based on Eq.~(\ref{eq:mr}), they would not significantly alter our results of the abundance of \acp{PBH} \cite{DeLuca:2020sae}.

\section{Limits on PBH abundance}\label{sec:re}

\begin{figure}
    \includegraphics[width =1. \columnwidth]{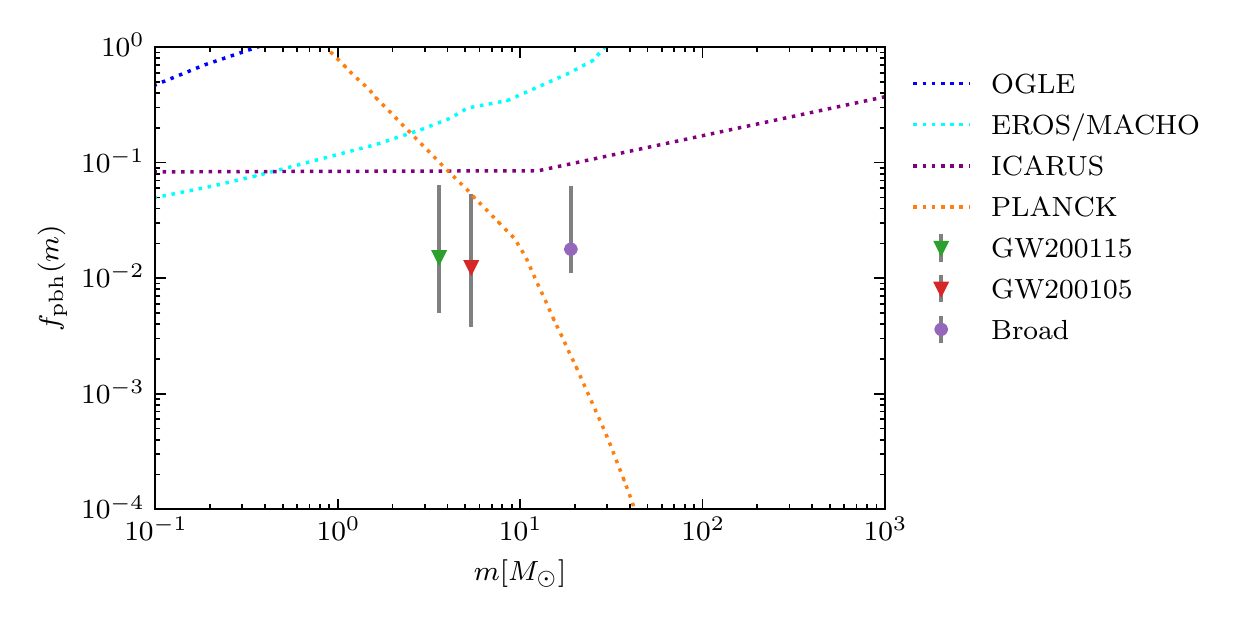}
    \caption{Constraints on $f_{\mathrm{pbh}}$ with $90\%$ CL error bars (gray solid) inferred from the merger rates for GW200105 (red triangle) and GW200115 (green triangle), and the broad mass function (purple circle). The existing upper limits on $f_{\mathrm{pbh}}$ at $90\%$ CL from OGLE (blue), EROS/MACHO (cyan), ICARUS (purple), and Planck (orange) are shown for comparison. }\label{fig:fpbh}
\end{figure}

Our results are shown in Fig.~\ref{fig:fpbh}. 
We depict the limits on $f_{\mathrm{pbh}}$ estimated from the merger rates of GW200105 (red triangle) and GW200115 (green triangle), and the broad mass function (purple circle), respectively. 
The gray solid lines denote the error bars at $90\%$ CL, due to uncertainties in the measurements of the merger rates \cite{LIGOScientific:2021qlt}. 
The dotted curves stand for the existing upper limits at $90\%$ CL from OGLE (blue) \cite{Niikura:2019kqi}, EROS/MACHO (cyan) \cite{EROS-2:2006ryy,Macho:2000nvd}, ICARUS (purple) \cite{Oguri:2017ock}, and Planck (orange) \cite{Serpico:2020ehh}, which are shown here for comparison. 

In the case of single \ac{PBH} event, the mass function is determined by the considered event, i.e. GW200105 or GW200115. 
Given the observational uncertainty, we derive the allowed abundance of \acp{PBH} which matches the observed merger rate of the event \cite{LIGOScientific:2021qlt}. 
We find that both of the events are well explained by such a scenario, when we take $f_{\mathrm{pbh}}\simeq\mathcal{O}(10^{-2})$ that is compatible with the existing best constraint $f_{\mathrm{pbh}}\simeq\mathcal{O}(10^{-1})$ from caustic crossing \cite{Oguri:2017ock}. 
We also find that the inferred upper limits are close to the existing upper limit, implying that such a scenario may be further tested in the near future. 
Furthermore, it is interesting to find that the mass function with $m_\ast=5.4M_\odot$ and $\sigma=0.54$ can almost precisely explain GW200105 and GW200115 simultaneously. 
But this coincidence may be accidental. 

In the case of the broad mass distribution, the mass function is determined by all the observed black hole binaries, since all of them are assumed to be primordial. 
This has been done by using maximum-likelihood analysis in an existing literature \cite{DeLuca:2020sae}. 
In our work, the merger rate is predicted via integration over $m_{i}\in [2.5,40]M_\odot$ and $m_{j}\in [1,3]M_\odot$, following a method consistent with Ref.~\cite{LIGOScientific:2021qlt}. 
Given $f_{\mathrm{pbh}}\simeq\mathcal{O}(10^{-2})$, we conclude that the merger rates predicted by this model are compatible with those of GW200105 and GW200115 reported by \ac{LVC} \cite{LIGOScientific:2021qlt}, although the predicted constraint on $f_{\mathrm{pbh}}$ is in tension with that inferred from the cosmic microwave background measured by Planck \cite{Serpico:2020ehh}. {In addition, by combining the broad mass function from Ref.~\cite{DeLuca:2020sae} with the two recently observed events, i.e. GW200105 and GW200115, we obtain a joint result of $m_\ast=16.4M_\odot$ and $\sigma=1.0$, which is slightly different with that of Ref.~\cite{DeLuca:2020sae}.}

\section{Conclusions}\label{sec:cd}

In this work, we have shown that the scenario of \ac{PBH} binary coalescences can account for the observed merger rates of GW200105 and GW200115. 
We computed the merger rates in two different models. 
The first one assumed that only GW200105 and GW200115 have a primordial origin. 
The second one assumed that all the observed black hole binaries are primordial. 
Given $f_{\mathrm{pbh}}\simeq\mathcal{O}(10^{-2})$, we found that both of the two models can explain the recently reported events.
For the former, the required abundance of \acp{PBH} were found to be compatible with the existing upper limits from other observations. 
The inferred upper limits were shown to be close to the existing upper limits, implying that such scenarios can be further tested by observations of \ac{SGWB} in the near future \cite{Wang:2016ana,Wang:2019kaf,Kohri:2020qqd,Mukherjee:2021ags,DeLuca:2021hde}. 
However, the latter was found to be in tension with the exiting observational limits. 
In addition, we found that the log-normal mass function of \acp{PBH} with $m_\ast=5.4M_\odot$ and $\sigma=0.54$ almost simultaneously explains the observed merger rates of GW200105 and GW200115, including the center values and error bars. 
In summary, depending on the abundance of \acp{PBH}, the scenario of \ac{PBH} binaries formed through the early-Universe channel could well explain GW200105 and GW200115 events. 

\begin{acknowledgements}
We would like to acknowledge Prof. Dr. Zhoujian Cao and Dr. Zucheng Chen for helpful discussions. 
S.W. is supported by the grants from the National Natural Science Foundation of China with Grant NO. 12175243, the Institute of High Energy Physics with Grant NO. Y954040101, the Key Research Program of the Chinese Academy of Sciences with Grant NO. XDPB15, and the science research grants from the China Manned Space Project with NO. CMS-CSST-2021-B01. 
Z.C.Z. is supported by the National Natural Science Foundation of China with Grant No. 12005016. 
\end{acknowledgements}

\bibliographystyle{unsrt}
\bibliography{pbh-nsbh}   


\end{document}